\documentclass[amsmath,amssymb,preprint]{revtex4}

\usepackage{graphicx}
\usepackage{dcolumn}
\usepackage{bm}

\begin{document}

\title{A stochastic interspecific competition model to predict the
behaviour of \emph{Listeria monocytogenes} in the fermentation
process of a traditional Sicilian salami}
\author{\normalsize Alessandro Giuffrida$^1$\footnote{e-mail: agiuffrida@unime.it},
Davide Valenti$^2$\footnote{e-mail: valentid@gip.dft.unipa.it},
Graziella Ziino$^1$, Bernardo Spagnolo$^2$, Antonio
Panebianco$^1$}\affiliation{\small $^1$Dipartimento di Sanit\`{a}
Pubblica Veterinaria, Section of Inspection of Food of Animal
Origin, Polo Universitario dell'Annunziata, Viale Annunziata, 98168
Messina,
Italy\\
$^2$Dipartimento di Fisica e Tecnologie Relative, Group of
Interdisciplinary Physics\footnote{Electronic address:
http://gip.dft.unipa.it}, Universit\`a di Palermo and CNISM-INFM\\
Viale delle Scienze, I-90128 Palermo, Italy}

\begin{abstract}
The present paper discusses the use of modified Lotka-Volterra
equations in order to stochastically simulate the behaviour of
\emph{Listeria monocytogenes} and Lactic Acid Bacteria (LAB) during
the fermentation period (168 h) of a typical Sicilian salami. For
this purpose, the differential equation system is set considering
$T$, $pH$ and $aw$ as stochastic variables. Each of them is governed
by dynamics that involve a deterministic linear decrease as a
function of the time t and an "additive noise" term which
instantaneously mimics the fluctuations of $T$, $pH$ and $aw$. The
choice of a suitable parameter accounting for the interaction of LAB
on \emph{L. monocytogenes} as well as the introduction of
appropriate noise levels allows to match the observed data, both for
the mean growth curves  and for the probability distribution of
\emph{L. monocytogenes} concentration at
168 h.
\end{abstract}

\maketitle
\thispagestyle{empty}

\section{Introduction}Predictive microbiology aims to develop accurate
and versatile mathematical models able to describe microbial
evolution in food products as a function of environmental
conditions. According to Whiting and Buchanan~\cite{Whiting},
predictive models are classified as primary, secondary and tertiary.
Primary models describe microbial evolution as a function of time.
Secondary models relate parameters which appear in primary modelling
to environmental conditions such as temperature ($T$), $pH$, water
activity ($aw$), etc. Tertiary models combine primary and secondary
models (e.g., the Pathogen Modelling Program, designed by the USDA
or the Growth Predictor developed by IFR) including, in more
extended versions, the possibility to import a T history in order to
predict remaining shelf life with regard to Specific Spoilage
Organisms, as in the Seafood Spoilage Predictor~\cite{Dalgaard}.
Modelling the microbial evolution as a function of fluctuating
environmental conditions (dynamic models) is very important for
practical predictive applications, especially when the technology of
some products, such as ripened meats and cheeses, is based on a
continuous modification of T, pH, relative humidity (RH), etc.
However, for these kinds of products, dynamic
models~\cite{Baranyi1,Bovill,Xanthiakos} can lead to overestimate
the real bacterial growth if they do not take into account the
inhibition by the competitive natural microflora. The bacterial
competition within the natural microflora is a complex issue in
modelling microbial evolution and it has been studied by several
authors~\cite{Pin,Gimenez,Vereecken,Van_Impe}. An interesting
approach to describe bacterial competition is based on the
generalized Lotka-Volterra model~\cite{Lotka,Volterra}. This
provides basic equations for the population growth of two
interacting species. A prototype model structure for mixed microbial
populations in food products was proposed by Dens et
al.~\cite{Dens}. This consists of a system of four differential
equations and combines the advantages of the Lotka-Volterra model
for two competing species with those of Baranyi \& Roberts'
model~\cite{Barany2} as a classical predictive growth model. Powell
et al.~\cite{Powell} used the above mentioned system of equations in
order to interpret some empirical results for \emph{Escherichia
coli} O157:H7 in ground beef, and they showed that the seemingly
incongruous data was consistent with the interspecific competition
model. Furthermore, the authors considered the effect of varying the
growth rates stochastically, introducing an uniform random
distribution for ?max of both species into the model; in this way,
the incorporation of microbial community dynamics into the food
safety risk assessment process was explored. The construction of
probabilistic predictive models, based on stochastic dynamics, is
very important for the relation between predictive microbiology and
quantitative risk assessment. An extensive review on predictive
microbiology~\cite{McMeekin} indicated that the use of stochastic
models produces remarkable effects, since it allows to move away
from the worst-case scenario. In this regard, Nauta~\cite{Nauta},
discussing the separation of uncertainty and variability in
quantitative microbial risk assessment models, proposed the
introduction of some stochastic terms into a system consisting of
primary and secondary growth models~\cite{Ratkowsky1}. A very
different approach to stochastically model bacterial growth and
inactivation is based on considering the microbial population
behaviour as the mean of the behaviour of many
cells~\cite{Baranyi3,Baranyi4,Swinnen}. In this case, measuring the
single cell growth or survival parameters, and repeating these
measurements for a wide number of cells, it would be possible to
express Lag-time and $\mu_{max}$ in terms of probability
distributions. This theoretical approach, further discussed in
Kutalik et al.~\cite{Kutalik}, was used by M\'{e}tris et
al.~\cite{Metris} for \emph{E. coli}, Francois et
al.~\cite{Francois1} and Francois et al.~\cite{Francois2} for L.
monocytogenes. In a recent paper Ponciano et al.~\cite{Ponciano}
presented a new application of a stochastic ecological model based
on the stochastic version of the well-known Verhulst logistic
differential equation. By using an extensive experimental data set
and requiring the specification of a likelihood function, this
stochastic model allows to analyse the influence both of
deterministic and random variations of the environmental conditions
on microbial growth dynamics. In light of the above trends in
predictive microbiology, the development of new models should
satisfy three important requirements: i) the application of a
dynamical approach, namely based on the use of differential
equations; ii) the presence of interactions among food microbial
communities; iii) the presence of stochastic terms in equations in
order to take into account the influence of random variations of the
environmental conditions. An interesting answer to the
aforementioned requirements could be provided by incorporating
stochastic growth rates into an interspecific competition model
described by the Lotka-Volterra equations. The dynamics of the
species can be affected by random fluctuations directly, through a
term of multiplicative noise in the species equations, or
indirectly, through a term of additive noise which mimics the
fluctuations of environmental conditions~\cite{Spagnolo1,Valenti1}.
The aim of this work is to propose a stochastic approach for
predictive microbiology. More specifically, we add noise terms in
the time evolution of three basic parameters T, pH, aw, in an
attempt to obtain a stochastic model for the bacterial growth
dynamics, starting from the above mentioned generalized
Lotka-Volterra equations. This approach was used in order to
stochastically simulate the behaviour of L. monocytogenes and Lactic
Acid Bacteria (LAB) during the fermentation step of a typical
Sicilian salami; the model was validated by taking into account the
data of a previous challenge test for L.
monocytogenes~\cite{Giuffrida}. We have chosen to consider only the
fermentation step since, as well known, the fluctuations of T, pH
and aw are wider in this phase than in the dry-curing, where T
especially remains constant. Furthermore, the aim of this work is to
evaluate L. monocytogenes behaviour under conditions of potential
growth and in the presence of LAB competitive activity while, during
the dry-curing in strict sense, the values of T, pH and aw fall
below the L. monocytogenes "growth region"~\cite{Le_Marc}.

\section{Materials and methods}
\subsection{Model}The stochastic interspecific competition model is
structured in order to simulate the behaviour of L. monocytogenes
and Lactic Acid Bacteria (LAB) during the fermentation step (7 days)
of a traditional Sicilian salami. According to the data of a
previous work [28], this is done considering T, pH, and aw linearly
decreasing respectively from 20°C to 12°C, from 5.8 to 5.6,
\cite{Giuffrida}and from 0.972 to 0.946, during 168 h. The system of
four differential equations proposed by Dens et al.~\cite{Dens} and
Powell et al.~\cite{Powell},
\begin{eqnarray}
\frac{dN_{Lmo}}{dt} &=& \mu_{max \thinspace Lmo}\thinspace
N_{Lmo}\frac{Q_{Lmo}}{Q_{Lmo}+1}\left(1-\frac{N_{Lmo}+\beta_{Lmo/LAB}\thinspace
N_{LAB}}{N_{max \thinspace Lmo}}\right) \label{LV1}\\
\frac{dQ_{Lmo}}{dt} &=& \mu_{max \thinspace Lmo}Q_{Lmo}\label{Q1}\\
\frac{dN_{LAB}}{dt} &=& \mu_{max \thinspace LAB}\thinspace
N_{LAB}\frac{Q_{LAB}}{Q_{LAB}+1}\left(1-\frac{N_{LAB}+\beta_{LAB/Lmo}\thinspace
N_{Lmo}}{N_{max \thinspace LAB}}\right) \label{LV2}\\
\frac{dQ_{LAB}}{dt} &=& \mu_{max \thinspace Lmo}Q_{LAB},\label{Q2}
\end{eqnarray}
was used as basic model. Here NLmo and NLAB are, respectively, the
population densities of L. monocytogenes and Lactic Acid Bacteria at
time t; $\beta_{max \thinspace Lmo}$ and $\beta_{max \thinspace
LAB}$ are the maximum specific growth of both species and $N_{max
\thinspace Lmo}$ and $N_{max \thinspace LAB}$ are the theoretically
maximum population densities of both species under monospecific
growth conditions. The coefficients $\beta_{Lmo/LAB}$ and
$\beta_{LAB/Lmo}$ are, respectively, the interspecific competition
parameters of LAB on L. monocytogenes and vice-versa. $Q_{Lmo}$ and
$Q_{LAB}$ represent, respectively, the physiological state of the
two species. Furthermore, according to Baranyi and
Roberts~\cite{Barany2}, we obtain the Lag-time
\begin{equation}
\lambda(t)=\frac{-\ln\alpha(t)}{\mu_{max}(t)} \label{lambda}\\
\end{equation}
with $\alpha(t)$ given by
\begin{eqnarray}
\alpha_{Lmo}(t) &=& \frac{Q_{Lmo}(t)}{1+Q_{Lmo}(t)}\label{alpha1}\\
\alpha_{LAB}(t) &=& \frac{Q_{LAB}(t)}{1+Q_{LAB}(t)}.\label{alpha2}
\end{eqnarray}
The below second order model Eq.~(\ref{mu_dalgaard}) (so-called
Ratkowsky or square root type model), developed by Tom Ross at the
University of Tasmania (UTAS model) and reported by  Gim\'{e}nez and
Dalgaard~\cite{Gimenez}, was incorporated into Eq.~(\ref{LV1}) while
T , pH and aw were time-dependent.
\begin{eqnarray}
\sqrt{\mu_{max \thinspace Lmo}} &=&
0.14776 \cdot (T-0.88)\cdot(1-e^{\thinspace 0.536 \thinspace (T-41.4)})\cdot \sqrt{Aw-0.923} \cdot \sqrt{1-10^{4.97-pH}}\nonumber\\
&\cdot& \sqrt{1-\frac{LAC}{3.79(1+10^{pH-3.86})}}\cdot(350-NIT)/350,
\label{mu_dalgaard}
\end{eqnarray}
where the values 0.88, 41.4, 0.923, 4.97 and 350 represent
respectively $T_{min}$ ($^\circ C$), $T{max}$ ($^\circ C$),
$aw_{min}$, $pH_{min}$ and $NIT_{max}$ (nitrite concentration in
ppm). LAC is the lactic acid concentration ($g \thinspace l^{-1}$)
and was obtained by using the following system of differential
equations, proposed by Leroy and De Vuyst~\cite{Leroy1} and Leroy et
al.~\cite{Leroy2}
\begin{eqnarray}
\frac{dLAC}{dt} &=& -YLAC_S \thinspace \frac{dS}{dt}\label{LAC}\\
\frac{dS}{dt} &=& \frac{-1}{YLAB_S \thinspace
\frac{dN_{LAB}}{dt}-m_S LAB}. \label{S}
\end{eqnarray}
Here YLACS is the yield coefficient for the production of lactic
acid from the fermentable sugar S ($g \thinspace l^{-1}$), YLABS and
mSLAB are two coefficients which express the depletion of
fermentable sugar $S$ as a function of the instantaneous
modifications of Lactic Acid Bacteria concentration ($dN_{LAB}/dt$;
see equations 1c and 1d). In order to solve the equation 1c we used
the below second order model proposed by Wijtzes et
al.~\cite{Wijtzes}
\begin{equation}
\mu_{max \thinspace LAB} = -0.00234 \cdot (aw-0.928) \cdot (pH-4.24)
\cdot (pH-9.53)\cdot(T-3.63)^2, \label{mulab}
\end{equation}
where the values 0.928, 4.24, 9.53 and 3.63 represent respectively
$aw_{min}$, $pH_{min}$, $pH_{max}$ and $T_{min}$. According to Leroy
and De Vuyst~\cite{Leroy1}, the term YLACS (Eq.~(\ref{LAC})) was set
to $1$, while the following models were used in Eq.~(\ref{S})
\begin{eqnarray}
YLAB &=& 3.3 \cdot (6.10^{-4} \thinspace T^2 - 0.044 \thinspace T +
1.03) \cdot (-0.13 \thinspace pH^2 +1.48 \thinspace pH -3.88) +
0.035 \qquad
\label{YLAB}\\
m_S LAB &=& 0.3 \cdot (-0.1583 \thinspace pH^2 + 1.98 \thinspace pH
- 5.59) \cdot YLAB^{-1}_S - 0.23. \qquad \label{S}
\end{eqnarray}
For the initial value of $Q_0$, the procedure of Baranyi et
al.~\cite{Baranyi1} was followed: a geometric mean value for the
physiological state parameter $\alpha(t)$ was estimated from several
growth curves of both bacterial populations, obtained from ComBase
(http://combase.arserrc.gov; IFR, Norwich, UK). The same procedure
was used in order to set the Nmax values, while the initial
bacterial values were those of a previous study~\cite{Giuffrida}
which were used to validate the present model (see the related
section). Finally, the system was solved, by numerical simulations,
to obtain predictions for the bacterial concentration during
time-dependent $T$ / $pH$ / $aw$ profiles.

\subsection{Noise and stochastic dynamics}

In the above section we have introduced a model of two interacting
species based on generalized Lotka-Volterra
equations~\cite{Lotka,Volterra,Valenti1}, where the growth rates of
the two species depend on $T$, $pH$ and $aw$. We assume that these
three parameters are driven by deterministic forces, which model the
external conditions imposed, for example, by the production
standards used in the food industry. However, real ecosystems
interact with a noisy nonstationary environment, so that parameters
such as $T$, $pH$ and $aw$, are also affected by random
fluctuations. To describe this continuous and noisy behaviour of
$T$, $pH$ and $aw$, we consider the following stochastic
differential equations~\cite{Gardiner}
\begin{eqnarray}
\frac{dT(t)}{dt} &=& k_T \thinspace t + \xi_T (t) \label{temperature}\\
\frac{dpH(t)}{dt} &=& k_{pH} \thinspace t + \xi_{pH} (t) \label{ph}\\
\frac{daw(t)}{dt} &=& k_{aw} \thinspace t + \xi_{aw}, \label{aw}
\end{eqnarray}
where the deterministic terms depend linearly on the time t and the
random terms $\xi_T (t)$, $\xi_{pH} (t)$, $\xi_{aw} (t)$ mimic the
fluctuations that affect $T$, $pH$, and $aw$, considering their
interaction with the environment. The coefficients $k_T$, $k_{pH}$,
$k_{aw}$ are the rates of $T$, $pH$ and $aw$, respectively.  $\xi_T
(t)$, $\xi_{pH} (t)$, $\xi_{aw} (t)$ are statistically independent
Gaussian white noises with zero mean and correlation functions
$<\xi_T (t)\xi_T (t')>=\sigma_T\thinspace\delta_T(t-t')$,
$<\xi_{pH}(t)\xi_{pH}(t')>=\sigma_{pH}\thinspace\delta_{pH}(t-t')$,
$<\xi_{aw}(t)\xi_{aw}(t')>=\sigma_{aw}\thinspace\delta_{aw}(t-t')$.
Here $\sigma_T$, $\sigma_{pH}$, $\sigma{aw}$ are the standard
deviations of the three Gaussian distributions, and they are the
intensities of the noise sources which affect $T$, $pH$, and $aw$.

\subsection{Scenarios}

Eqs.~(\ref{LV1})-(\ref{Q2}) were solved numerically, obtaining the
time series of the species concentrations for different scenarios,
that correspond to different values of the noise intensities and
interaction parameters.
\begin{table}[htbp]
\begin{center}
\includegraphics[width=16cm]{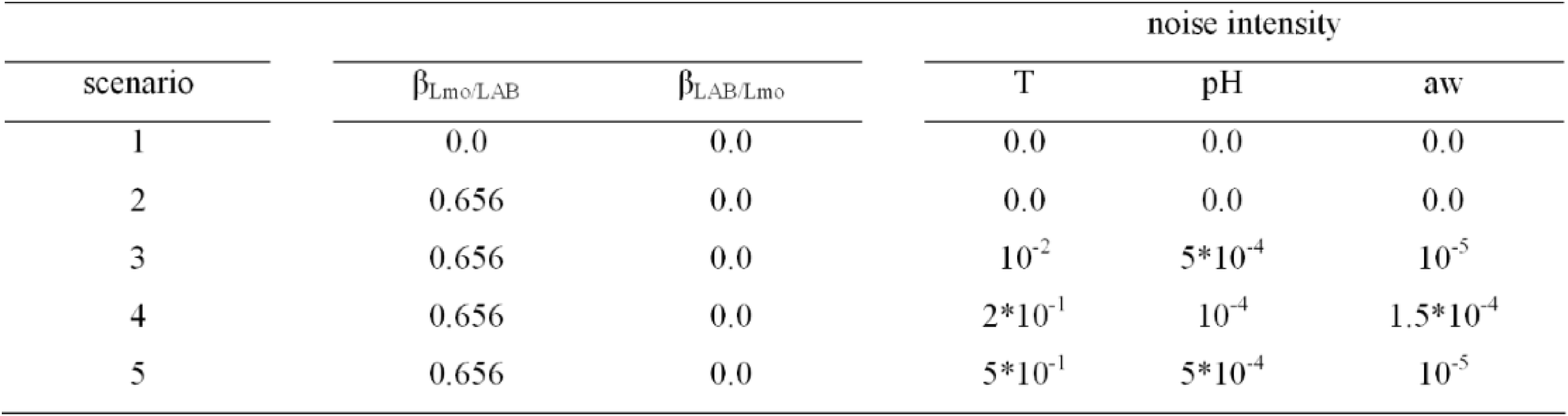}
\end{center}
\vskip-0.3cm \caption{\small Competition model parameters used for
each scenario. Scenario 1 assumes the absence of interaction between
species and a zero noise intensity. The "zero intensity" of noise is
maintained in the scenario 2, but the interaction term of LAB on
\emph{L. monocytogenes} is introduced. In scenario 3, a low
intensity of noise is introduced for $T$, $pH$ and $aw$  so that the
fluctuations of the three environmental variables affect weakly the
growth parameter $\mu_{max \thinspace Lmo}$. In scenarios 4 and 5
the noise intensity progressively increases whereas the interaction
terms remain constant.\bigskip}
 \label{table1}
 \vskip-0.3cm
\end{table}
Table~\ref{table1} summarizes the Lotka-Volterra competition model
parameters for the scenarios considered in this paper. For every
scenario we solved
Eqs.~(\ref{LV1})-(\ref{Q2}),~(\ref{lambda}),~(\ref{LAC})-(\ref{S})
by numerical integration, performing $1000$ realizations. The
initial conditions for the two specie concentrations were chosen, in
each realization, by setting a Gaussian distribution with mean and
standard deviation equal to those experimentally
observed~\cite{Giuffrida}. In particular, as a mean value and
standard deviation we used, respectively, Log $2.492$ cfu/g and
$0.170$ for \emph{L. monocytogenes}, and Log $6.825$ cfu/g and
$0.688$ for LAB. Scenario 1, used as a control scenario, assumes the
absence of interaction between species and a zero noise intensity.
In this case the system works like the logistic model of Baranyi et
al.~\cite{Barany2} in the presence of decreasing environmental
conditions ($T$ from 20 $^\circ C$ to 12 $^\circ C$, $pH$ from $5.8$
to $5.6$, and $aw$ from $0.972$ to $0.946$). The "zero intensity" of
noise is maintained in scenario 2, but introducing the interaction
term of LAB on \emph{L. monocytogenes} ($\beta_{Lmo/LAB} = 0.656$).
In scenario 3, a low intensity of noise is introduced for $T$ ($T =
10^{-2}$), $pH$ ($pH = 5 \cdot 10^{-4}$) and $aw$ ($aw = 10^{-5}$)
so that the fluctuations of the three environmental variables ($T$,
$pH$, $aw$) weakly affect the growth parameter $\mu_{max \thinspace
Lmo}$. In scenarios 4 and 5 the noise intensity progressively
increases whereas the interaction terms remain constant. Note that,
in scenarios 2, 3, 4, 5 (see table 1), the absence of any
competitive effects of \emph{L. monocytogenes} on LAB
($\beta_{LAB/Lmo} = 0$), implies $\beta_{Lmo/LAB} < N_{max
\thinspace Lmo}/N_{max \thinspace LAB}$, as a coexistence condition.
The value of $\beta_{Lmo/LAB}$ (see table 1) and those fixed for the
maximum population densities, $N_{max \thinspace Lmo} = 7.5$ and
$N_{max \thinspace LAB} = 9.3$, satisfy the previous inequality, so
that a coexistence regime for \emph{L. monocytogenes} and LAB is
established~\cite{Dens,Spagnolo1,Valenti1}. This agrees with the
empirical data on the L. monocytogenes behaviour during the
seasoning of salami
~\cite{Giuffrida,Campanini,Trussel,Villani,Nissen,Tyopponen}. The
nitrite concentration was set constantly at 90 ppm, according to the
data of a previous work~\cite{Giuffrida}.

\subsection{Validation of the model}

In order to validate the deterministic and stochastic predicted
results, we considered the experimental data obtained in a previous
study [28] in a challenge test for L. monocytogenes in a typical
Sicilian salami. In particular, the data referred to 63 samples of
S. Angelo Salami PGI (Protected Geographical Indication; EU
regulation 510/06) inoculated, after the stuffing into a natural
pork casing, according to the indication of Scott et
al.~\cite{Scott}, with a suspension (5 ml; Log $4,320$ cfu/ml) of a
strain of L. monocytogenes previously isolated in the same kind of
salami. The experimental contamination was performed by inoculating
the suspension of L. monocytogenes into 20 different sites of each
sample. Products, with an average weight $500\pm 10$g and diameter
of 50-60 mm, were coarse-grained and contained 3\% of NaCl, 100 ppm
of nitrate and 90 ppm of nitrite, according to the related product
specification. Before the stuffing, a commercial mixture of
Lactobacillus plantarum and Staphylococcus xylosus, for one group of
samples, and a mixture of Pediococcus acidilactici, for a second
group, were used. In the present study, however, the two groups were
considered together, since, as previously reported~\cite{Giuffrida},
the evolution of LAB and L. monocytogenes as well as $pH$, $aw$ and
weight loss were very similar. Nine products for each group were
analysed in triplicates at 0, 48, 120 and 168 hours after stuffing,
during the fermentation step (7 days), with regard to the
determination of L. monocytogenes count, Lactic Acid Bacteria, pH
and aw. The environmental temperature and RH were monitored by a
data logger (FT-102 Econorma, Treviso - Italy), in order to check
the fermentation programme characterized by a gradual decrease of
temperature from 20 $^\circ C$ to 12 $^\circ C$ and a gradual
increase of RH from $63$\% to $70$\%, within the considered time
interval (168 h). In this study we performed experiments using a
single inoculum level (Log $2.492$ cfu/g). This condition does not
affect the validity of our analysis since we expect that the
inoculation level plays a marginal role in the bacterial dynamics
when the system is far from strongly stressed conditions and the
initial concentration is greater than $1$ Log cfu/g~\cite{Besse}.
Therefore, we compared experimental and theoretical results by using
a single inoculation level over 54 trials and one profile of $T$,
$pH$ and $aw$ to obtain a statistical distribution for the
concentration at different times. In particular, the \emph{L.
monocytogenes} predicted mean growth curve was compared to the
observed one and the differences were statistically analysed through
the Root Mean Squared Error (RMSE). Furthermore, the chi-squared
test was used in order to evaluate the differences among the
probability distributions of the observed and predicted \emph{L.
monocytogenes} concentrations at 168 hours. It is worth noting that
the above study, based on a single experiment with 54 trials, was
not considered in view of a complete validation of the model, since
the aim of the work was  to show the role of environmental noise on
bacterial dynamics. For this purpose, we took into account a data
set where the variability is strongly reduced, in order to highlight
the effects of the random fluctuations of the environmental
parameters, through a comparison between observed data and
theoretical results. However, a validation of our model, in view of
a general application to the dynamics of two bacterial competing
species, needs wider investigation and it will be the subject of a
forthcoming paper.

\section{Results and discussion}

\subsection{Deterministic behaviour (scenarios 1-2)}

\begin{figure}[htbp]
\begin{center}
\includegraphics[width=13cm]{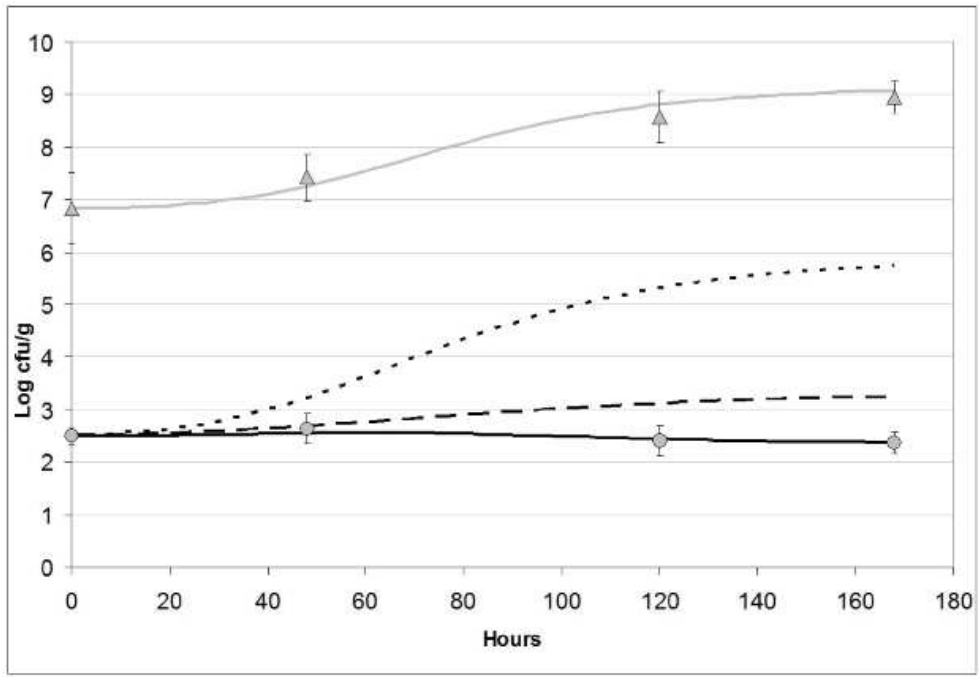}
\end{center}
\vskip-0.3cm \caption{\small Observed mean bacterial behaviour
(\emph{L. monocytogenes}: circles; LAB: triangles). Error bars
indicate $\pm$ 1 standard  deviation. Black line with little dashes
and black solid line represent the predicted values of \emph{L.
monocytogenes} concentration for scenarios 1 and 2, respectively.
Gray line indicates the predicted LAB growth for both scenarios. The
three curves are obtained with nitrite concentration set at 90 ppm.
The black line with large dashes shows the prediction of L.
monocytogenes concentration for scenario 1 with nitrite
concentration at 150 ppm.\bigskip}
 \label{fig1}
 \vskip-0.3cm
\end{figure}

The observed mean behaviour of \emph{L. monocytogenes}
(Fig.~\ref{fig1}) is characterized by a slight increase and a
subsequent decrease until a final (168 hours) concentration given by
Log $2.370$ $\pm$ $0.214$ cfu/g. This is very close to the initial
(0 hours) value, Log $2.492$ $\pm$ $0.170$ cfu/g. The prediction
carried out by using the parameters of Scenario 1 (interaction terms
and noise intensity set to 0) provides a behaviour which is very
different from the experimental data.
In Fig.~\ref{fig1} the black line with little dashes represents the
time behaviour of the \emph{L. monocytogenes} concentration for
scenario 1. Here, the maximum predicted value of the \emph{L.
monocytogenes} concentration is lower (Log $5.724$ cfu/g) than the
theoretical Maximum Population Density $N_{max \thinspace Lmo}$ (Log
$7.5$ cfu/g), and closer to the observed value. However, the
predicted behaviour of \emph{L. monocytogenes} for scenario 1 is
very different (RMSE = 2.2386) from the observed data (circles in
Fig.~\ref{fig1}). This indicates that the effect of environmental
hurdles, such as the lactic acid concentration modelled according to
equations 5a-b, cannot explain the complexity of the considered
microbiological system. A better prediction is obtained for scenario
1 by setting the nitrite concentration at 150 ppm (instead of 90 ppm
used in this work, according to the previous work~\cite{Giuffrida}).
However, also in this case the predicted growth (black line with
large dashes in Fig.~\ref{fig1}) does not agree with the observed
behaviour (RMSE = 0.5634). In scenario 2, according to Powell et
al.~\cite{Powell}, the interspecific interaction is considered. We
find that a suitable value of $\beta_{Lmo/LAB}$ exists for which the
mean \emph{L. monocytogenes} behaviour (black solid line in
Fig.~\ref{fig1}) is very close to the observed one (RMSE = 0.0449).
This result shows the fundamental role played by the interspecific
competition in view of investigating the bacterial dynamics of
\emph{L. monocytogenes} and LAB communities. The Lotka-Volterra
approach is able to simulate the competition between two
populations, describing different situations such as the mutual
interaction, the reduction of only one population ("low or no
growth") or the decline. These effects are related to the bacterial
concentration which, however, depends only on environmental
conditions since the interaction parameters, $\beta_{Lmo/LAB}$, do
not affect the maximum growth rate. The approach based on the
"Jameson effect" hypothesis~\cite{Besse} gives a phenomenological
description of \emph{L. monocytogenes} behaviour in real food, that
is, in the presence of other bacterial species, i.e. competitors. In
fact, according to the "Jameson effect" hypothesis, due to the
bacterial interaction, a different value of Nmax is measured,
without providing a "dynamical" explanation for this new value.
Conversely, previously measuring the $N_{max}$ value for \emph{L.
monocytogenes} in monoculture allows us to acquire knowledge of the
"free" (in the absence of bacterial competition) behaviour of
\emph{L. monocytogenes}, and to obtain the effect of the competition
by introducing an interaction term. The comparison between
experimental data, obtained in competition regime, and theoretical
results, calculated by using, in the competition model, the growth
parameters previously obtained from monoculture experiments,
indicates what the effect of the bacterial interaction is. At the
same time this allows to determine the values of the interaction
parameters, $\beta_{Lmo/LAB}$ and $\beta_{LAB/Lmo}$, for which the
theoretical distributions are in good agreement with the
experimental ones. Note that, in scenario 1, the model works as a
conventional predictive system (e.g. Baranyi and Roberts model)
where the bacterial behaviour is only governed by the three
parameters $Q$, $\mu_{max}$, $N_{max}$. In particular, $Q$ and
$\mu_{max}$ are related to environmental characteristics through a
secondary predictive model which derives from a monoculture set of
experimental data, while the limiting term $N_{max}$ is a static
parameter. In scenario 2, the conventional predictive approach
describes the gradual transition of the system to the Lotka-Volterra
dynamics. In fact, according to
Eqs.~(\ref{LV1},\ref{Q1},\ref{LV2},\ref{Q2}), the interaction terms,
which depend on the bacterial concentrations, begin to express the
competitive effect of LAB on \emph{L. monocytogenes} when the system
leaves the Lag-phase. This approach differs from that of Leroy et
al.~\cite{Leroy3} which studied the competitive interaction of
Lactobacillus sakei on \emph{L. monocytogenes}. According to their
previous studies~\cite{Leroy1,Leroy2} they modelled the production
and the activity of the bacteriocin by using mono- and co-culture
in-vitro conditions, showing a non-constant activity of
bacteriocins. In view of this aspect and taking into account the
real complexity of the considered food system, characterized by the
heterogeneity of LAB and presence of other interaction mechanisms
different from the bacteriocin production~\cite{Giraffa}, we have
chosen to describe the interaction mechanism of LAB on \emph{L.
monocytogenes} by using a single term ($\beta_{Lmo/LAB}$).

\subsection{Stochastic behaviour (scenarios 3-5)}

\begin{figure}[htbp]
\begin{center}
\includegraphics[width=13cm]{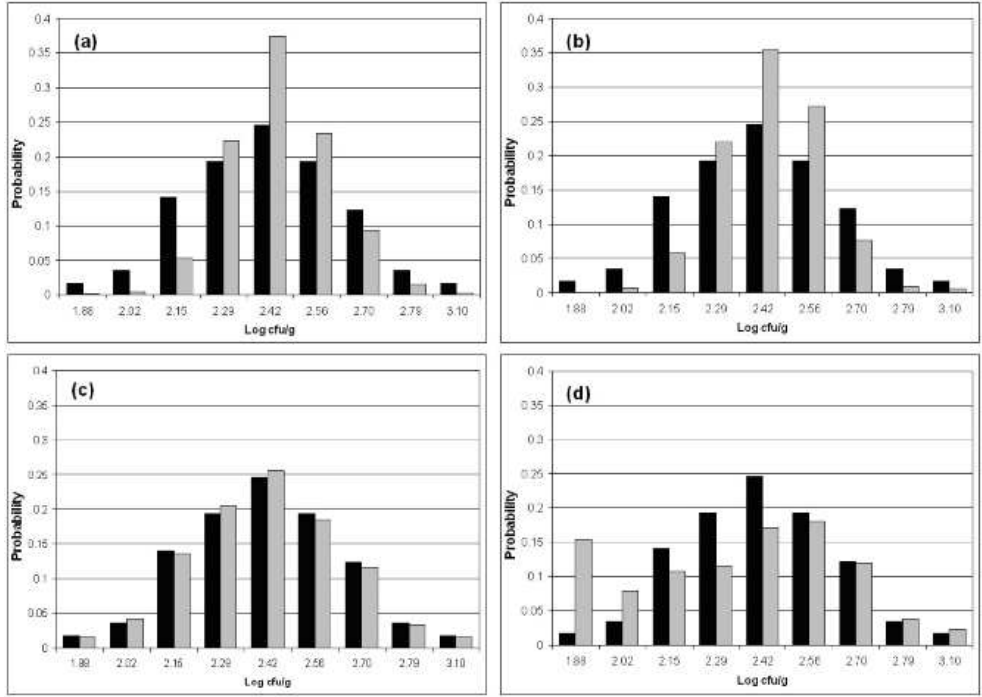}
\end{center}
\vskip-0.3cm \caption{\small . Observed (black bars) and predicted
(gray bars) probability distributions of \emph{L. monocytogenes}
concentration at 168 hours for scenarios 2 (panel a), 3 (panel b), 4
(panel c) and 5 (panel d). The total number of trials in the
experimental work is 54. In the theoretical approach we performed
$1000$ iterations.\bigskip}
 \label{fig2}
 \vskip-0.3cm
\end{figure}

In Fig.~\ref{fig2} we show the probability distribution of \emph{L.
monocytogenes} concentration at 168 hours for scenario 2 (panel a),
scenario 3 (panel b), scenario 4 (panel c) and scenario 5 (panel d),
compared to the observed probability distributions.
\begin{table}[htbp]
\begin{center}
\includegraphics[width=16cm]{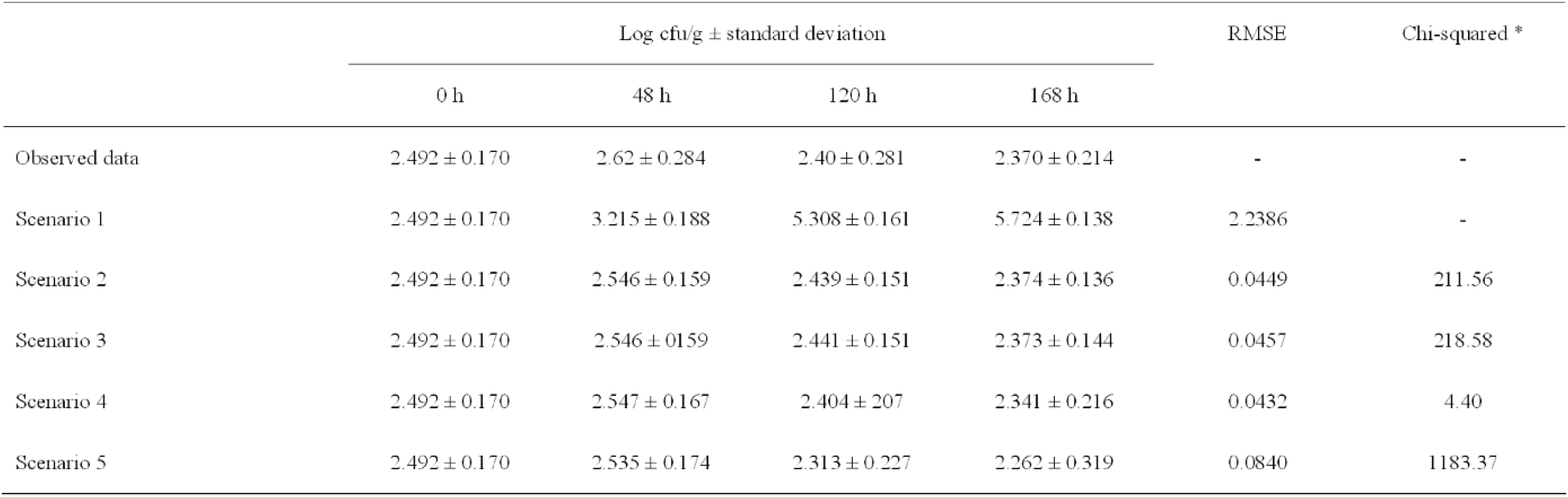}
\end{center}
\vskip-0.3cm \caption{\small . Observed and predicted \emph{L.
monocytogenes} mean values at 0, 48, 120 and168 hours, for each
scenario. The Root Mean Squared Error (RMSE) indicates the agreement
of the mean predicted curves to the observed ones in each scenario.
The chi-squared values are referred to the \emph{L. monocytogenes}
probability distribution at 168 h for scenarios 2, 3, 4 and
5.\bigskip}
 \label{table2}
 \vskip-0.3cm
\end{table}
Table~\ref{table2} reports the results for the chi-squared test,
RMSE and mean value of the \emph{L. monocytogenes} concentration for
all scenarios. As Table 2 and Fig.~\ref{fig2} show, the increase of
noise produces a reduction of the mean value of the L. monocytogenes
concentration at 168 h. In particular, for scenarios 2 and 3 (0 or
low noise intensities) the central part of the predicted probability
distribution takes on values significantly larger than the observed
one (panels a and b of Fig.~\ref{fig2}). For higher levels of noise
the values of the theoretical probability distribution are reduced,
and the predicted and observed probability distributions are in very
good agreement (Fig.~\ref{fig2}c). A further increase of the noise
intensity causes the predicted probability distribution to become
very different from the observed one (Fig.~\ref{fig2}d). Moreover,
concerning the RMSE values, as shown in Table 2, the noise intensity
of scenario 4 allows an enhancement of predictive model performances
to be obtained, producing the lowest RMSE value. Conversely, a lower
noise intensity (scenario 3), as well as a higher one (scenario 5),
produces a reduced fitting of the \emph{L. monocytogenes} mean
concentration with the observed data (Table~\ref{table2}).
Furthermore, in scenarios 2 and 3, the simulation provides a
distribution characterized by a small standard deviation and
pronounced symmetry around the central value. On the other hand, in
scenario 5, the data distribution (with a high standard deviation)
shows a marked peak at the minimum value. Therefore, the overall
evaluation of these results, obtained from stochastic dynamics,
shows that the environmental noise causes a global effect consisting
of a reduction of the \emph{L. monocytogenes} mean value
concentration. However, at the same time, a suitable level of noise
intensity allows to obtain bacterial growth values, whose
probability distribution matches the observed one very well. This is
the most relevant result of the present study. It is important to
stress that we are not interested in measuring the intensity of the
noise that affects the experimental data. However, environmental
parameters, i.e. $T$, $pH$ and $aw$, undergo random fluctuations,
always present in "open systems" such as that considered in this
work. Here, we intend to show that the observed data distributions,
can not be reproduced by the proposed model in the absence of noise
(scenario 2). Conversely, a suitable level of noise allows to obtain
theoretical results in good agreement with experimental findings.
This aspect could play a key role in view of incorporating
stochastic microbial predictive models (such as the proposed one)
into a risk assessment process, since the introduction of the
appropriate level of noise can influence the precision in the
expression of the probabilistic "output" related to the
concentration of a foodborne disease agent. In our study, for
example, the observed percentage of samples with a \emph{L.
monocytogenes} concentration at h 168 $\leq$ Log $2.0$ cfu/g
(regulatory critical limits in EU) was 4.4\%, while the predicted
percentages in scenarios 3, 4 and 5, were, respectively, 0.4\%,
5.2\% and 22\%. Note that the separation between uncertainty and
variability, which is usually fundamental in the application of
stochastic models~\cite{Nauta,Pouillot}, was not introduced in the
present study, since the observed data are obtained by a single
strain of \emph{L. monocytogenes}. In general, the bacterial cells,
obtained from the same strain, should exhibit the same average
biological and physiological properties. Therefore, our
inoculations, and then the corresponding concentrations, belong to
the same statistical distribution. This means that the
hyperparameters, i.e. expected value and standard deviation, are the
same for the initial concentrations used in the experimental
trials~\cite{Pouillot}. This condition suggests the absence of
variability in the initial conditions over the different trials,
while the uncertainty, which is connected with the difficulty to
inoculate all salami at precisely the same concentration and the
technical uncertainty such as the error of the enumeration method,
was expressed by a Gaussian distribution with mean and standard
deviation equal to the experimentally observed ones (Log $2.492\pm
0.170$ cfu/g for \emph{L. monocytogenes}). However, the variability
could be considered by changing the growth rate parameter for each
different strain in our Lotka-Volterra stochastic model. Another
interesting consideration regards the enhancement of predictive
model performances obtained by increasing the noise until the level
of scenario 4 (RMSE values, Tab. 2). From a biological point of
view, this accounts for the environmental noise, that is, random
fluctuations of external variables such as $T$. The presence of
noise influences the growth rate and, indirectly, the interaction
between bacterial species. It is important to recall that in many
fields where population dynamics is studied, the noise effects on
ecological systems are the subject of an intensive
investigation~\cite{Gallagher,Zimmer,Bjornstad}. Theoretical
analyses and experimental results have showed that noise, in the
presence of nonlinear dynamics, is responsible for several
counterintuitive phenomena, such as stochastic resonance
~\cite{Benzi1,Benzi2,Jung1,Jung2,Mantegna1,Lanzara,Gammaitoni,Alley,Mantegna2},
noise enhanced
stability~\cite{Mantegna3,Mantegna4,Agudov1,Agudov2,Fiasconaro}, and
noise delayed extinction~\cite{Valenti1,Spagnolo2}, which are not
present in purely deterministic regimes. Therefore, noise and its
effects have become a well established subject in physics,
chemistry, and biology~\cite{Freund}. The contemporary presence of
noise and nonlinear interactions in ecological systems is
responsible for the appearance of a rich dynamics, which corresponds
to the real complexity of natural systems. From a theoretical point
of view, this situation can be described by using a model where both
the internal nonlinear interactions of the system and the noisy
interaction with the environment are taken into account, giving rise
to a complex behaviour of the system, which is very sensitive to
initial conditions, various deterministic external perturbations and
random fluctuations always present in nature. This paper presents a
further evolution of the interspecific competition model proposed by
Dens et al.~\cite{Dens} and Powell et al.~\cite{Powell}, in order to
reproduce the complexity of some food systems during their
production. In this regard, taking into account the generalized
Lotka-Volterra equations, we introduced $T$, $pH$ and $aw$ as
stochastic variables. The dynamics of $T$, $pH$, $aw$, affecting
bacterial rate, obey stochastic differential equations that involve
both a deterministic term, varying linearly as a function of time,
and a random term, that is responsible, at each time t, for
fluctuations (noisy behaviour) of $T$, $pH$, $aw$. Initially, we
considered the deterministic decrease and increase, respectively, of
$T$ and $RH$ in the seasoning rooms and the bacterial metabolic
activity (e.g. sugar fermentation). Moreover, the $T$ and $RH$ of
seasoning rooms can be also affected by random fluctuations, beside
the decrease imposed by production standards. In this way, in
principle, each single cell composing a food bacterial population
interacts with a different environment at each time t, having a
different Lag-time and growth rate. The introduction of stochastic
terms, expressed by equations~\ref{temperature},~\ref{ph},~\ref{aw},
into the Lotka-Volterra equations reproduces the presumable noisy
behaviour which affects the Lag-time, the growth rate and therefore
the bacterial concentration proportionally to the noise intensity,
allowing the transition from a deterministic to a stochastic
predictive model. The probabilistic model used in this work has some
similarities with that of Francois et al.~\cite{Francois1} and
Francois et al.~\cite{Francois2} for \emph{L. monocytogenes}, but
the approach methodology is very different. In fact, the above
mentioned authors measured empirically the fluctuations of the
species concentrations at constant environmental conditions
(individual-based approach), obtaining a statistical distribution by
fitting the different data sets. In our case, the data distribution
is the consequence of environmental noise, modelled as a white
Gaussian-distributed noise, whose SD represents the intensity.
Therefore, while the data distribution obtained with the
"individual-based approach" accounts for intra-specific differences
in adaptation to environmental parameters, our approach mainly
considers different cell behaviour as a consequence of environmental
heterogeneity. As this study shows, the real behaviour of \emph{L.
monocytogenes} in meat products during the fermentation step is
mainly affected by bacterial interactions which are, however,
dependent on environmental fluctuations too. Moreover, other
implications could be explored using, for example, a time correlated
noise as theoretically suggested by Spagnolo et al.~\cite{Spagnolo1}
or introducing a "multiplicative
noise"~\cite{Spagnolo1,Valenti2,Ciuchi}, which directly affects
specie concentrations. A further development of the stochastic
dynamical model proposed in this paper could also consist of using a
growth/no-growth term~\cite{Le_Marc,Ratkowsky2,Valero}. This implies
that fluctuating environmental conditions could cause the growth
rate to go below a given threshold (no-growth region), contributing
to the generation of a richer dynamics: both growth and no-growth
cells of a bacterial population could have a non-vanishing
probability to appear at the same time. Therefore, the model, in a
probabilistic sense, should be predictive: it could allow to
calculate which is the probability that, at a certain time t, the
bacterial growth (or no-growth) takes a given value. In conclusion,
our approach as well as all its further developments could be useful
for the incorporation of the predictive microbiology models into the
quantitative risk assessment process.

\end{document}